\documentclass[journal]{IEEEtran}
\usepackage{graphics}
\usepackage[colorlinks=false, urlcolor=blue, pdfborder={0 0 0}]{hyperref}

\usepackage{amsmath}
\usepackage{caption}
\usepackage{subcaption}
\usepackage{amssymb}
\usepackage{multirow}

\usepackage{cite}

\usepackage{array}

\usepackage{xfrac}
\usepackage{xcolor}

\ifCLASSINFOpdf
\else
\fi
\begin{document}
%
\title{Coop-RPL: A Cooperative Approach to RPL-based Routing in Smart Grid AMI Networks}
%
%
%

%

\author{
        Zahra~Aslani and
        Adnan~Aijaz
\thanks{Z. Aslani is with the Department of Electrical, Computer \& IT Engineering, Qazvin Branch, Islamic Azad University, Qazvin, Iran.

A. Aijaz was with the Centre for Telecommunications Research, King's College London, London, UK.

}}

\maketitle


\begin{abstract}

The next generation of electric grid has a scalable and reliable bi-directional communication infrastructure known as Advanced Metering Infrastructure (AMI) to allow monitoring and controlling of grid resources. In this kind of communication system, comprising thousands of resource-constrained nodes such as smart meters, the routing protocol plays a critical role to guarantee reliability and low latency for data delivery. The Routing Protocol for Low Power and Lossy Networks (RPL) currently is standardized by IETF which is expected to meet the requirements of AMI networks as the standard routing protocol. On the other hand, cooperative routing has gained a lot of attention recently. Cooperative routing improves performance over traditional routing by exploiting the broadcasting nature of wireless channels. Our objective in this paper is to propose a cooperative approach to RPL for application in AMI networks. Our proposed protocol, termed as Coop-RPL, is specially designed for the challenges of AMI networks. 
Coop-RPL ensures reliable packet transmission by selecting an optimal relay node. Performance evaluation demonstrates the efficiency of Coop-RPL for AMI networks with the aim of achieving increased PDR and reduced number of retransmission and end-to-end delay.

\end{abstract}


%
\IEEEpeerreviewmaketitle

\section{Introduction}
%
%
%
%

\IEEEPARstart{R}{ecent} studies on electric power systems reveal that the conventional power grid is not only inefficient but also unsuitable  to meet the ever increasing power demands. The next generation of the electric grid, called smart grid, is a fast developing technology that leads to efficiency, sustainability, reliability and adaptability through incorporation of advanced bi-directional communications, automated control, and distributed computing capabilities \cite{TowardSmartGrid}. One of the key elements of smart grid is  the Advanced Metering Infrastructure (AMI), which provides two-way communication between smart meters and data concentrators. The whole system is linked to a Meter Data Management System (MDMS) which analyzes the data collected from the smart meters and manages it to facilitate different applications \cite{CommunicationNetwork}.

In AMI networks, the design challenges include reliable and and low-latency data delivery. In such networks, fading and interference effects make the wireless link between smart meters unstable. Hence,  proper routing functionalities become critical in meeting the requirements of reliability and low-latency for data delivery. Recently, an IPv6 routing protocols has been standardized by the IETF which is known as Routing Protocol for Low Power and Lossy Networks (RPL) \cite{IPv6Routing}. RPL aims to overcome routing challenges in resource constrained environments and intends to support a variety of applications such as industrial monitoring, building automation, connected homes, health care, environmental monitoring, urban sensor networks, asset tracking, etc. \cite{RPLNutshell}. RPL is expected to the de-facto routing protocol for practical AMI networks. 

A recent study indicates that under dense AMI deployments, RPL might select sub-optimal paths with highly unreliable links \cite{RPLRouting}. This is because RPL does not have complete knowledge of link qualities and normally it measures the quality of a link only when data is sent through that link. The reliability of communications between smart meters is the key factor in any AMI application. Therefore, this issue becomes more critical. To overcome this limitation, the route construction and data delivery processes in RPL require a revisit. 

On the other hand, cooperative communications in wireless networks has gained a lot of attention recently. A number of studies have been conducted in literature on integrating cooperative communications with routing protocols. By exploiting the broadcast nature of wireless channels, a neighboring node can act as a relay to forward overheard packets to the intended receiver. Therefore, the cooperative routing can provide performance gains compared to traditional routing protocols \cite{CooperativeCommunications}, \cite{DistributedRouting}, \cite{CooperativeRouting}. Such features of cooperative communication paradigm are particularly attractive for AMI networks.

In this paper, we explore a cooperative communication approach to improve the reliability of AMI networks. We integrate cooperative communications with RPL to design a novel routing protocol for practical AMI networks. RPL nodes may suffer from unreliability due to the fact that RPL might select suboptimal paths. Hence, finding a relay node with higher link quality will improve the reliability by decreasing the number of re-transmissions and end-to-end delay, and enhancing the packet delivery ratio (PDR). In our proposed protocol, termed as Coop-RPL, some nodes will be chosen as relay nodes based on the specified parameters to find the high quality links and facilitate reliable transport of data to the concentrator node. These parameters are considered in separate routing classes to observe the effects of each of them in more details. Then we study Signal to Interference plus Noise Ratio (SINR), the number of active connections, child nodes, and Expected Transmission Count (ETX) as selected parameters in three routing classes which aim to improve PDR and reduce number of re-transmission and end-to-end delay. Therefore, a node with the highest priority will be chosen as the selected relay node to cooperatively forward data through the default parent based on the aforementioned parameters.

The rest of the paper is organized as follows. We begin our discussion by providing an overview of RPL. This is followed by a review of some of the most important RPL-based
routing solutions for AMI networks. After that we describe the framework of our proposed protocol, Coop-RPL. Performance evaluation and comparison with state-of-the-art
protocols has been conducted. Finally, the paper is concluded and some future work directions identified.

\section{Overview of RPL}
%
%
%
%

RPL is designed by the IETF ROLL working group to facilitate the connectivity in low power and lossy networks. RPL utilizes one or more \textit{Directed Acyclic Graphs} (DAGs) in order to maintain network state information. A DAG is a directed graph with no cycle and the destination nodes act as the root of it. In the case of the AMI networks, the DAG root typically is the gateway node. Each node is associated with a rank value to maintain its position and determine its relation with other nodes in the DAG. The rank of nodes monotonically increases in a downward direction. A node rank is only comparable with those nodes that have similar or smaller ranks in comparison with the selected node rank to avoid any routing cycle. To build a DAG, the gateway node will advertise a control message so-called \textit{DAG Information Object} (DIO). Nodes in the range of the gateway node receive the DIO message and decide to join the DAG or not. Once a node tends to join the DAG, it adds the sender of the DIO message to its parent list and computes its own rank based on an Objective Function (OF). After this process, the node transmits the DIO message with the updated rank information to its neighbors. The objective function defines how a node calculates the rank based on the rank of its neighbors and selects the default parent node within its parent list. Since DIO messages are frequently issued throughout the DAG, a joined node may receive more than one DIO message. In such a situation, the joined node should select to process the received DIO message or discard it. As a result of this processing, if the node evaluates the DIO message, it can remain in its current position or can update and improve its position to a lower rank according to the objective function. Otherwise, the joined node could ignore the DIO message and keep its previous position in the existing DAG. 
After the DAG is formed, each node is capable of forwarding any upward traffic (destined to the gateway) through its parents to the DAG root. If a node does not receive a DIO message, it will send a \textit{DAG Information Solicitation} (DIS) control message to ask the neighbors for transmitting a DIO message.

In addition, the RPL defines a strategy for downward traffics from the gateway to a node. In this case, a node creates and sends a unicast \textit{Destination Advertisement Object} (DAO) control message to its parent as a unicast packet to specify the reverse route information. The intermediate nodes are able to record the reverse path information from the DAO message and as a result, a complete route is created between the root and the sender node.

Each node periodically generates DIO messages using a \textit{trickle timer} to maintain the DODAG. In fact, the trickle timer technique optimizes the message transmission, while keeping the network conditions stable. In a stable condition, the trickle time will increase but if an inconsistent condition is observed in the network, RPL will reset the trickle timer to a minimum value.
Further details on RPL can be found in references \cite{RPLNutshell} and \cite{CriticalEvaluation}.

\section{Existing RPL-based routing Solutions for AMI networks}

%
%
%

Recent studies mainly focus on the existing RPL-based routing solutions and their performance evaluation in AMI networks.  In \cite{RPLBased}, Wang \emph{et al.} offered a detailed performance of RPL with some modifications for reliable and low latency data delivery in inward and outward traffic for AMI networks. The authors proposed a novel ETX based rank computation method which was used for DAG structure and maintenance.  The proposed protocol provides high end-to-end reliability for the inward unicast traffic in AMI networks. Besides, the authors explore a reverse path mechanism for the outward unicast traffic by processing  inward data traffic. 

Similarly in \cite{RPLRouting}, Ancillotti \emph{et al.} investigated the performance of RPL and analyzed the route-level attributes, such as path stretch, route lifetimes, dominance, and flapping RPL in a typical AMI system, focusing on routing stability and reliability performance. Their studies might facilitate the design of new mechanisms to improve the reliability and adaptability of RPL. For example, they investigated more sophisticated hysteresis techniques that mitigate the risk of using low-quality links. Furthermore, lightweight channel probing techniques could be integrated into RPL to improve routing efficiency. The authors also explore how routing reliability can be improved with the use of more advanced routing schemes such as multi-path and network coding. 

In \cite{OpportunisticRPL}, Gormus \emph{et al.} described a practical mesh networking solution derived from extensions proposed to the routing protocol for RPL to realize automated metering communications. The authors develop an efficient cooperative anycasting approach for wireless mesh networks to improve the efficiency of data transport. It has been shown that by using anycasting in protocol design, it was possible to improve  scalability and  utilization of the network. 

Thulasiraman \emph{et al.} \cite{RPLMultigateway} mainly focused on the performance evaluation of RPL protocol in mesh based multi-gateway AMI networks. The authors considered  multiple DODAGs with different gateway. They also proposed a new objective function for rank computation based on signal interference. Results show that this objective function outperforms ETX based rank computation. 

Complementary work has been done by Parnian \emph{et al.} In their recent work  \cite{RPLRoutingProtocol}, the role of two different objective functions (ETX and hop count) and their impact on performance of RPL routing in AMI networks was investigated. They evaluated two objective functions implementation.

Cognitive and Opportunistic RPL (CORPL) was developed by Aijaz \emph{et al.} \cite{CORPL} with the primary objective of enhancing RPL for cognitive radio enabled AMI networks. An opportunistic forwarding approach has been adopted to facilitate the requirements of the secondary network (cognitive AMI network) while protecting the primary users. Results show that CORPL not only improves the reliability and latency of data delivery but also reduces  interference to primary users.

\section{Coop-RPL Framework}
%
%
%
%
In this section, we introduce the framework of our enhanced RPL protocol in AMI networks. The aim of Coop-RPL is to use the concept of DAG-based approach of RPL with some modifications that are specifically suitable for AMI networks 
We consider a static multi-hop wireless AMI network which includes smart meter nodes and one gateway node (meter concentrator).

In order to retain the DAG structure of RPL in our approach, the same procedure is followed for Coop-RPL as explained earlier in section II. The gateway node transmits a DIO message. The \textit{Expected Transmission Count} (ETX) is used as the default link metric to calculate the rank of nodes. For instance, the ETX measurement of a link between node $a$ and node $b$ has been described by the equation $E_{ab}=\sfrac{1}{p_{ab}}$ where $p_{ab}$ is the probability that node $b$ receives a successful transmission from node $a$. 

The ETX of each link will be evaluated when the link starts to transmit data traffic and the results will be updated simultaneously. 
In this way, the node with the lowest rank will be selected as the default parent.
It should be considered that each node in Coop-RPL has a default parent (similar to RPL) which is selected based on the ETX evaluation.
Therefore, the route from the sender (child) node to the default parent is considered as the direct path.

In Coop-RPL protocol, we assume that before sending data through the default parent, it is possible to use a relay node or not. A relay node is selected from a set of nodes called candidate relay nodes. These nodes are neighbors of the sender node which can be selected as the optimal relay to cooperatively forward data to the next node. 

In the first stage, so-called \textit{Filtering phase}, the neighbor nodes of the sender node which are not good enough in terms of the rank will be removed. We assume that relay nodes are part of the DAG, so there is no need to recalculate their ranks. In this section, we just select the neighbor nodes which have less rank than the sender node.
In the second stage, based on some parameters, we describe the procedure of finding the eligible neighbor nodes from the first stage which can be selected as candidate relay nodes. After that, the best one will be chosen among them. Coop-RPL considers three routing classes that in each of them, we propose a parameter to find the qualified relays:\\
\\- \textit{SINR\ (Class A)},\\
- \textit{Traffic\ and\ Number\ of\ Children\ (Class B)},\\
- \textit{ETX\ (Class C)}\\

\begin{figure}
\centering
\scalebox{.3}{\includegraphics{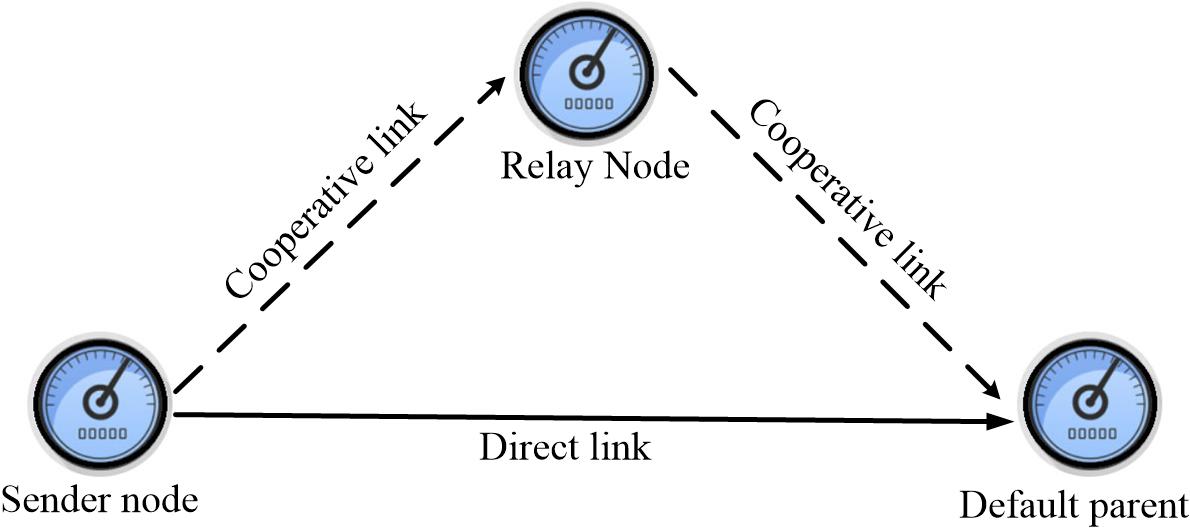}} \\
\scalebox{.45}{\includegraphics{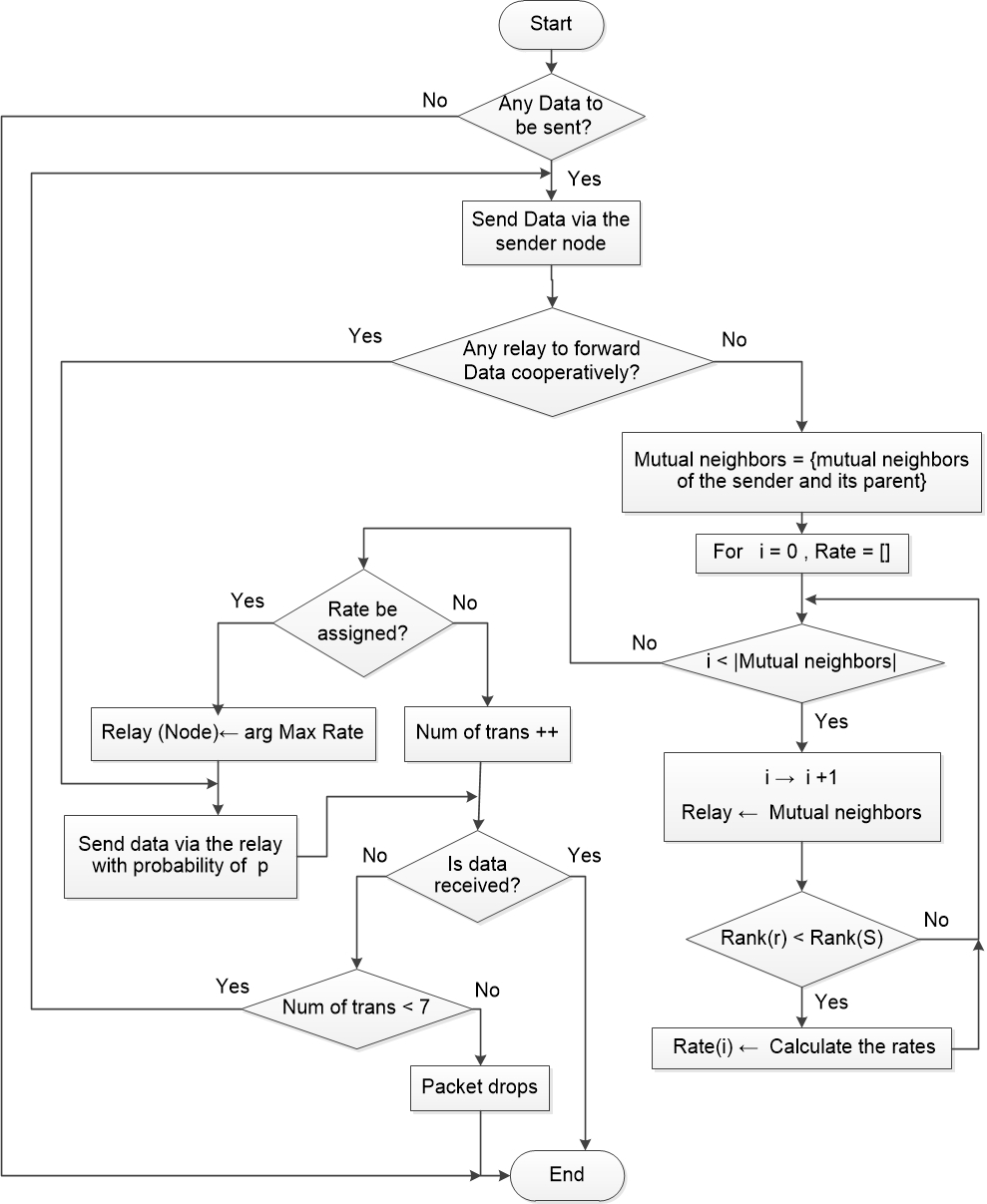}}
\caption{An illustration of cooperative transmission along with the flowchart of the proposed protocol.}
\label{fig:10}
\end{figure}

\textbf {\textit{Class A:}} Considering a cooperative transmission as shown in Fig.\ref{fig:10}, our method selects neighbor nodes as candidate relay nodes which are eligible to be chosen as the optimal relay node, if the quality of SINR in both cooperative links (source to relay and relay to destination) is better than the direct link (source to destination). The condition is given by
\begin{align}
\label{eqn_3}
& \text{if}\ SINR_{S,r} > SINR_{S,D}\ \text{\&\&} \ SINR_{r,D} > SINR_{S,D}, \\ \nonumber
& \text{then\ select\ $r$\ as\ a\ candidate\ relay\ node}
\end{align}
Where $SINR_{S,r}$ is received SINR at relay node $r$, $SINR_{r,D}$ is received SINR at node $D$ from the cooperative link and $SINR_{S,D}$ is received SINR  at node $D$ from the direct link.\\


\textbf{\textit{Class B:}} The important point in this class is considering the best path with the least traffic among all possible paths. Also, it is important to select a node as an optimal relay with the smallest number of children among others in order to avoid congestion in the cooperative path. Therefore, traffic will be distributed among more than one route.
Hence, the nodes with smaller number of children in contrast with the sender node will be selected as the candidate relay nodes. In addition, in order to estimate the data traffic, the number of active connections is considered. The number of active connections shows the total times a node is placed in the middle of the path as an intermediate node where the data traffic is forwarded through the gateway. Therefore, a node with smaller number of active connections compared to the sender node could be selected as the selected relay. The condition is given by
\begin{align}
\label{eqn_4}
& \text{if}\ NAC_{r} <NAC_{S}\
\ \&\& \ \ NCh_{r} < NCh_{S},\ \text{then}\\ \nonumber
& \text{select\ $r$\ as\ a\ candidate\ relay\ node}
\end{align} 
Where $NAC_{r}$ and $NAC_{S}$ show the number of active connections that relay node $r$ and sender node $S$ participate in respectively, and $NCh_{r}$ and $NCh_{S}$ show the number of relay node and sender node's children respectively.\\

\textbf{\textit{Class C:}} In this class, ETX is used as the default metric for relay nodes selection. 
A node $r$ will be considered as the candidate relay node, if the cooperative links $(S,r)$ and $(r,D)$ provide smaller ETX than the direct link $(S,D)$. This condition is given as follows.
\begin{align}
\label{eqn_5}
& \text{if}\ ETX_{S,D} > ETX_{S,r} + ETX_{r,D},\ \text{then}\ \\ \nonumber
& \text{select $r$ as\ a\ candidate\ relay\ node}
\end{align}
Where $ETX_{S,D}$ is related to the ETX of the direct link $(S,D)$ and the ETX of the cooperative link $(S,r)$ is defined by $ETX_{S,r}$ and finally $ETX_{r,D}$ represents the ETX of the cooperative link $(r,D)$.\\

In the third stage, coefficients (weights) are given to each of the mentioned parameters in order to calculate a value called \textit{Rate} for each candidate relay node. We should mention that in addition to the rank, the rate is considered for each candidate relay node based on the following formula to choose the best relay node among them.
\begin{align}
\label{eqn_1}
Rate&=W_{SINR}\times min\ ({SINR}_{S,r},{SINR}_{r,D}) \\ \nonumber
&-W_{traffic}\times Traffic_r \\ \nonumber
&-W_{NCh}\times Ch_r \\ \nonumber
&-W_{ETX}\times ({ETX}_{S,r}+{ETX}_{r,D})
\end{align}
$W_{SINR}, W_{Traffic}, W_{NCh},$ and $W_{ETX}$ are design parameters such that $W_{SINR}+W_{Traffic}+W_{NCh}+W_{ETX}=1$.
For \textit{Class A}, $W_{SINR}\gg W_{Traffic}, W_{NCh}, W_{ETX}$,
which means that a candidate relay node with the highest SINR (best quality of cooperative links) could be chosen as the relay node. Based on the formula of rate, since the SINR of cooperative links (sender to relay and relay to default parent) is not equal, we consider the minimum one to calculate the rate of each candidate relay node.
In \textit{Class B}, $W_{Traffic}, W_{NCh} \gg W_{SINR}, W_{ETX}$, 
hence, the number of children and data traffic play the most significant role in this class. While the number of children and traffic have negative effects on the results, the negative forms of these parameters have been considered in the rate formula.
Finally, for \textit{Class C}, $W_{ETX} \gg W_{SINR}, W_{Traffic}, W_{NCh}$,
therefore ETX is the most important parameter in this class. Since links with lower ETX have higher priorities, like \textit{class B}, we have considered the negative form of ETX. It means that a candidate relay node with a lower value of ETX in its cooperative links has a higher chance to be chosen as the selected relay.

Then we select the best relay with maximum rate as the \textit{“Selected Relay”}.
\newcommand{\argmin}{\arg\!\max}
\begin{equation*}
\label{eqn_2}
Selected\ Relay= \arg\max_i (Rate_i)
\end{equation*}

After considering the selected relay node, we may use the selected relay node with a probability to cooperatively send data through the default parent. 
We have introduced a sub-option to the DIO message to enable reporting of the node's relay to other nodes. The procedure of using a relay node to cooperatively forward a data packet or not and also the criteria of selecting the appropriate neighbor nodes as candidate relay nodes are shown in Fig. \ref{fig:10}.

\section{Performance Evaluation}
\begin{figure}
  \centering
    \scalebox{.4}{\includegraphics{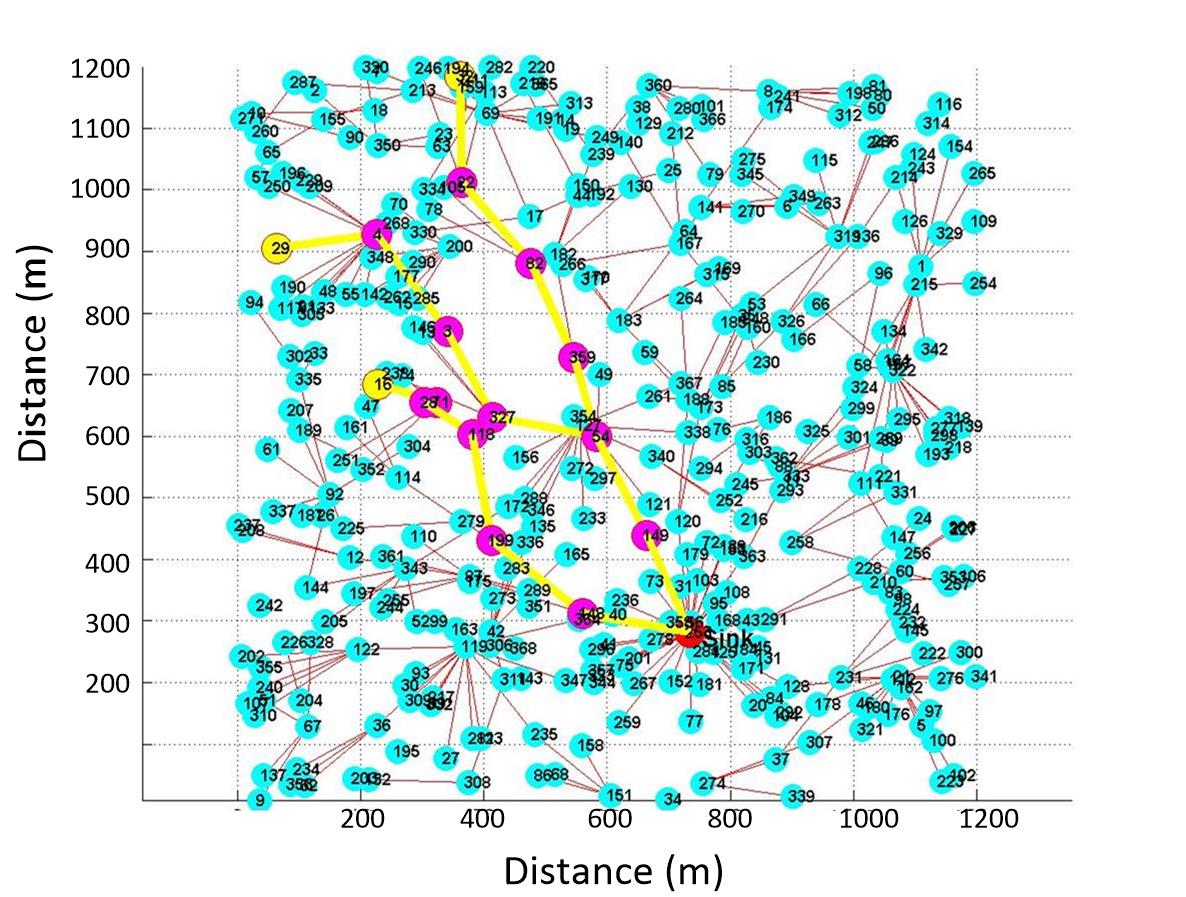}}
  \caption{Simulated network topology. Nodes are connected in the form of a DAG where numbers show node IDs.}
  \label{fig:2}
\end{figure}

In this section, we conduct a performance evaluation of Coop-RPL via simulations. A square region of side 1200 meters is considered within which nodes are Poisson  distributed. The channel model between a pair of smart meter nodes accounts for large-scale path loss and small-scale Rayleigh fading. Each node transmits at 2W and has a transmission range of 10 meters. The trickle timer is set to 100 ms.  An instance of simulated topology is shown in  Fig.\ref{fig:2}


\begin{figure}
  \centering
    \scalebox{.4}{\includegraphics{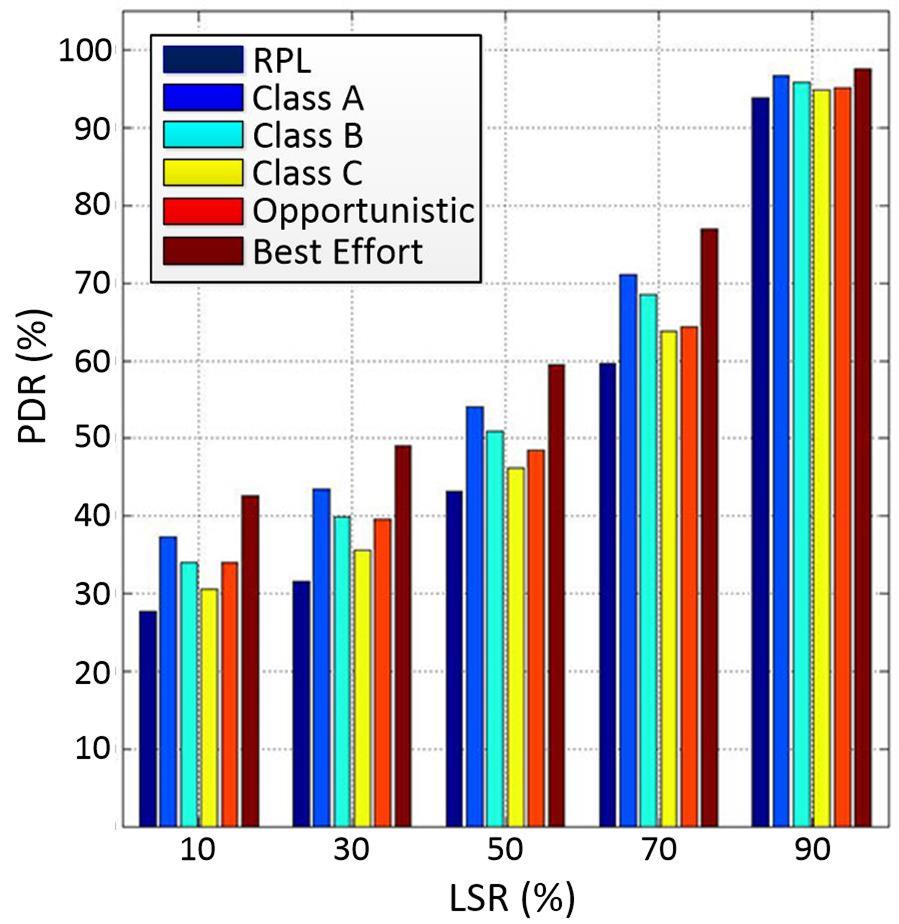}}
  \caption{PDR performance comparison vs. Link Success Rate for different protocols}
  \label{fig:3}
\end{figure}

\begin{figure}
  \centering
    \scalebox{.4}{\includegraphics{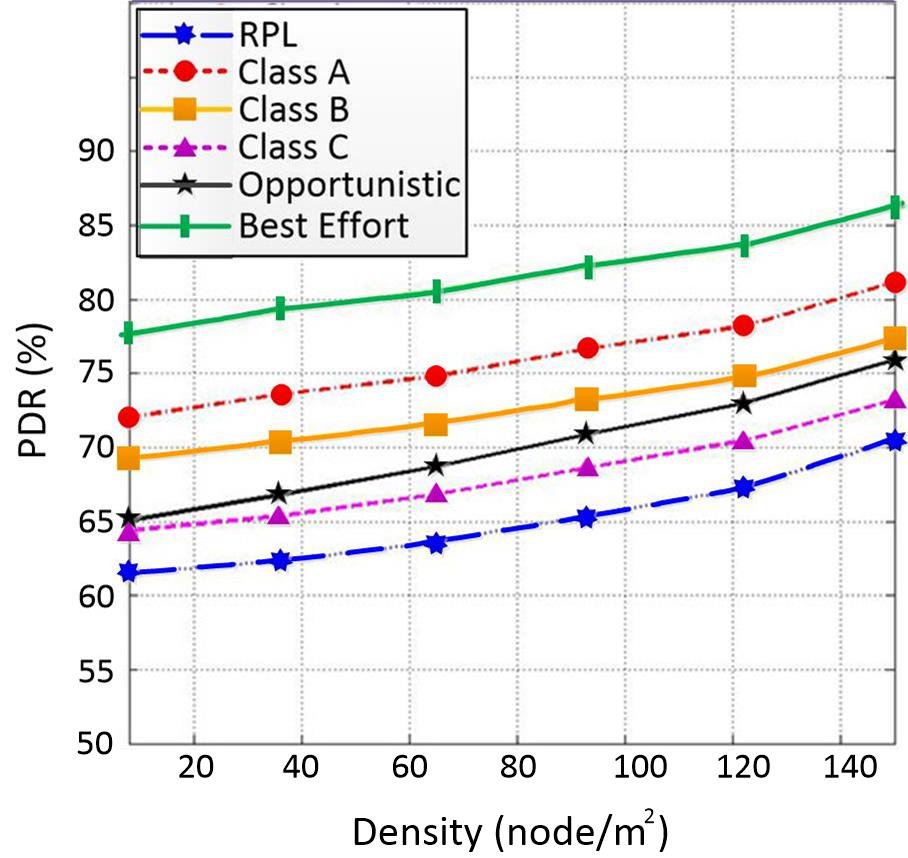}}
  \caption{PDR performance comparison vs. Density for different protocols}
  \label{fig:4}
\end{figure}

We evaluate the performance of Coop-RPL under the network scenario described above and compare it with RPL and opportunistic RPL. We use three nodes in the forwarding set for opportunistic RPL in our simulation. Fig. \ref{fig:3} and Fig. \ref{fig:4} show the plots of the packet delivery ratio against the link success rate and density of the network respectively. The ratio of the number of received packets to the total number of transmitted packets represents the performance evaluation in terms of Packet Delivery Ratio (PDR). The PDR records the fraction of packets sent by different nodes that are actually delivered to the gateway. Also 1000 packets are generated from different nodes and the average PDR is calculated relative to the Link Success Rate (LSR) and density ratio. As it can be seen in Fig. \ref{fig:3}, the PDR increases when the probability of link success rises. Also Fig. \ref{fig:4} shows that the PDR improves with increasing the density of the network. This is mainly due to the improved channel with the next hop owing to reduction in path loss. SINR is the dominant parameter in \emph{Class A} and has the highest value in comparison with other classes in Coop-RPL, RPL and opportunistic RPL protocol. So, it seems that this parameter has the most important effect on selecting the best relay node in contrast with other parameters in these two figures. Besides, in this class we consider wireless interference in path calculation, while ETX in RPL protocol does not calculate signal interference. Also, it is evident from Fig. \ref{fig:3} and Fig. \ref{fig:4} that Coop-RPL outperforms RPL in all classes and opportunistic RPL in \emph{Class A} and \emph{Class B}. In addition, these figures show that the Coop-RPL utilizes the diversity of routes and hence improves the PDR by reducing the number of re-transmissions.
In best-effort case, we consider coefficients of all parameters and as it is clear in Fig. \ref{fig:3} and Fig.\ref{fig:4}, it achieves the highest PDR in comparison with RPL and opportunistic RPL protocol.

\begin{figure}
  \centering
    \scalebox{.4}{\includegraphics{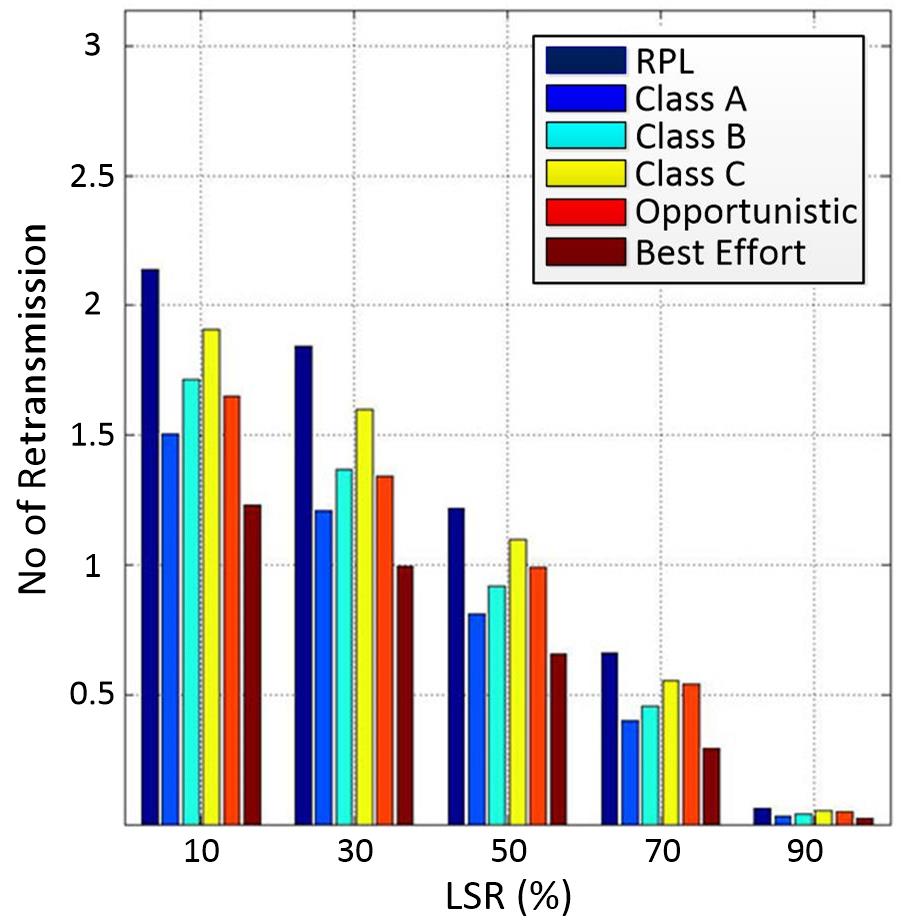}}
  \caption{Number of re-transmission vs. Link Success Rate in different protocols}
  \label{fig:7}
\end{figure}

Fig. \ref{fig:7} shows the plot of the number of re-transmission against the link success rate. It is concluded from this figure that by increasing the ratio of the LSR, the number of re-transmission significantly decreases due to the successful delivery of data. This figure indicates that Coop-RPL has better performance mainly because the relay node may try to send the same packet through the default parent, if a transmission is unsuccessful in Coop-RPL. This should reduce the need for re-transmission compared to the standard RPL, where unsuccessful transmissions will lead to re-transmissions. In addition, in RPL protocol, if after multiple attempts the packet drops, it will fail to receive to the gateway and consequently the packet will be lost. 
It is clear from Fig. \ref{fig:7} that \emph{Class A} has the smallest number of re-transmission among other classes in Coop-RPL, standard RPL, and opportunistic RPL protocol. So, this class selects the path with the highest quality that causes the smallest number of re-transmission. Also, as it can be seen in Fig. \ref{fig:7}, by considering coefficients of all parameters in the form of best effort in Coop-RPL, we achieve the smallest number of re-transmission in contrast with RPL and opportunistic RPL.

\begin{figure}
  \centering
    \scalebox{.4}{\includegraphics{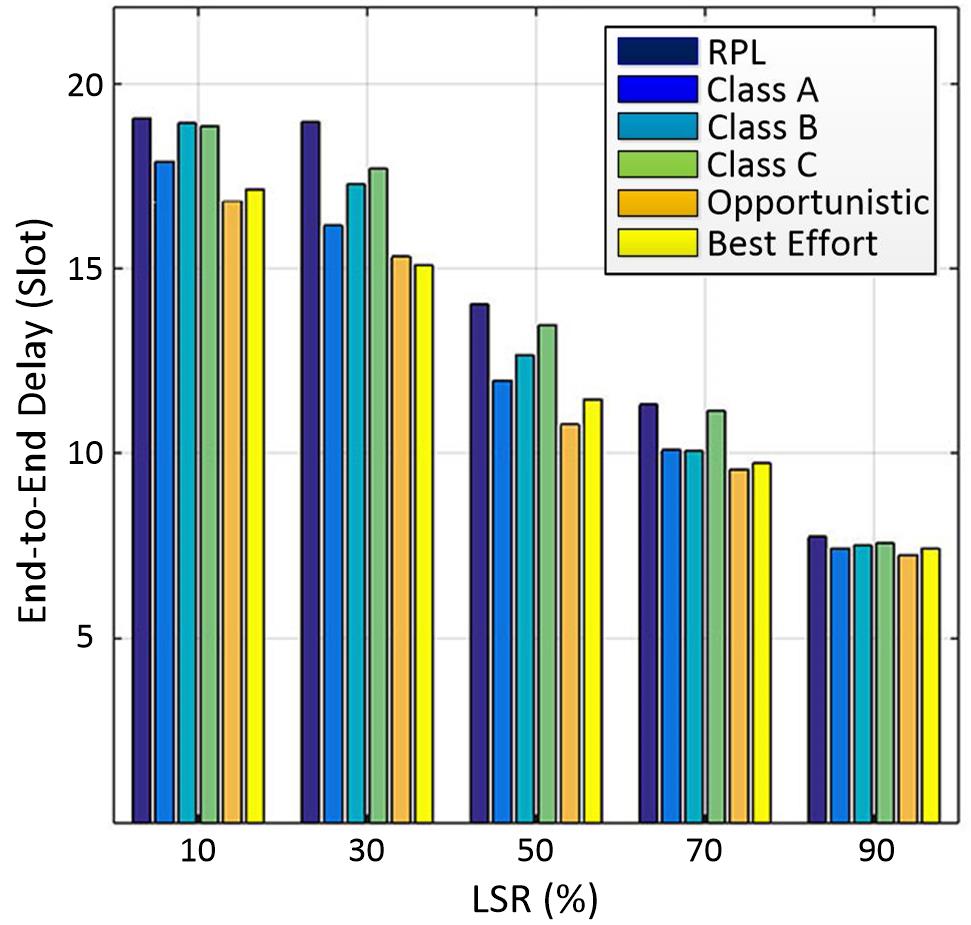}}
  \caption{End-to-End delay comparison vs. Link Success Rate for different protocols}
  \label{fig:9}
\end{figure}

Finally, in Fig. \ref{fig:9} the average end-to-end delay performance of proposed protocol is evaluated against the link success rate. As is shown in Fig. \ref{fig:9}, the increase in the LSR reduces the average end-to-end delay in the network. End-to-end delay is defined as the average time to successfully transmit a packet from a node to the root in DODAG. Fig. \ref{fig:9} shows that the less number of re-transmission provided by the cooperative links is able to significantly reduce the end-to-end delay. The Coop-RPL classes have less delay in comparison with RPL protocol, because it uses a relay node to cooperatively forward data through the sink. Also, this figure demonstrates that \emph{Class A} of Coop-RPL has the least delay where the optimal relay node is selected based on SINR parameter. Therefore, it is apparent from this figure that the relay node provides a lower end-to-end delay in good channel conditions. In all, our results indicate that under the described network scenario, Coop-RPL protocol produces a satisfactory performance in contrast with RPL and opportunistic RPL protocol in AMI networks.

\section{Conclusion}
In this paper, we have developed a novel method of enhanced RPL based routing protocol for AMI networks which is called Coop-RPL. Our results show a significant improvement both in the reliability and low latency data delivery of different application. This was previously considered as an essential challenge in AMI networks. We present the Coop-RPL results in three classes of cooperative forwarding approach that not only improve the reliability of AMI networks, but also reduce the end-to-end delay. In these classes, by selecting an eligible relay node, we achieve higher PDR and reduce the number of re-transmission. Therefore, Coop-RPL suggests an effective and viable solution for practical AMI networks. We also evaluate and compare the performance of Coop-RPL with other relevant protocols. The results show that Coop-RPL enhances PDR by up to 20\% and 10\% in best effort state compared to RPL and opportunistic RPL respectively. Besides, it reduces the end-to-end delay by up to 15\% compared to RPL protocol.


%





\ifCLASSOPTIONcaptionsoff
  \newpage
\fi

\end{document}